\documentclass[11pt]{article}
\usepackage{pdflscape}
\usepackage{multirow}
\usepackage{amsfonts}
\usepackage{amsmath}
\usepackage[pdftex]{graphicx}
\usepackage{longtable}
\usepackage{lscape}
\usepackage{pdflscape}
\usepackage{booktabs}
\usepackage{geometry}
\usepackage[toc,title,page]{appendix}
\PassOptionsToPackage{hyphens}{url}\usepackage{hyperref}
\usepackage[english]{babel}
\usepackage[utf8]{inputenc}
\usepackage{graphicx,kantlipsum}
\usepackage{graphicx}
\usepackage{etoolbox}
\usepackage{lipsum,caption,graphicx,bbm}
\usepackage{multicol}
\usepackage{caption}
\usepackage{subcaption}
\usepackage{breakcites}
\usepackage{comment}
\usepackage[flushleft]{threeparttable}
\usepackage[dvipsnames, table]{xcolor}
\usepackage{xcolor-solarized}
\usepackage{placeins} 

\usepackage{natbib}
\setcitestyle{authoryear,open={(},close={)}} 

\usepackage{rotating}
\usepackage{float}

\begin{document}

\title{Credit and Voting}
\author{Eleonora \textsc{BRANDIMARTI}\thanks{%
			Harvard Business School} \hspace{0.25cm} Giacomo \textsc{DE GIORGI}\thanks{%
			Institute of Economics and Econometrics-GSEM, University of Geneva BREAD, CEPR, IPA. Acknowledgments: De Giorgi acknowledges financial support through the SNF grant 100018\_182243.} \hspace{0.25cm}\\\medskip Jeremy \textsc{LAURENT-LUCCHETTI}\thanks{%
			Institute of Economics and Econometrics-GSEM, University of Geneva}}
\date{June 28, 2024 \\\bigskip  \vspace{-1cm}}

\maketitle


\begin{abstract}
There is a tight connection between credit access and voting. We show that uncertainty in access to credit pushes voters toward more conservative candidates in US elections. Using a 1\% sample of the US population with valid credit reports, we relate access to credit to voting outcomes in all county-by-congressional districts over the period 2004-2016. Specifically, we construct exogenous measures of uncertainty to credit access, i.e. credit score values around which individual total credit amount jumps the most (e.g. around which uncertainty on access to credit is the highest). We then show that a 10pp increase in the share of marginal voters located just around these thresholds increases  republican votes by 2.7pp, and reduces that of democrats by 2.6pp. Furthermore,  winning candidates in more uncertain constituencies tend to follow a more conservative rhetoric.
\end{abstract}

\newpage

\section{Introduction}
\label{sec:intro}

The opportunity for all US citizens to climb the social ladder and pursue the "American dream" is a central tenet of the US social contract. This requires that upward economic mobility remains accessible and widely visible.  Access to credit and home ownership are considered pillars of socioeconomic  mobility as they allow for wealth accumulation, e.g. by building equity and reducing housing costs. Recent research highlights that these advantages further translate into higher inter-generational income mobility  \citep{HERKENHOFF2021345}. 
Despite a majority of Americans aspiring to home ownership, they face several common hurdles, such as difficulties in putting together a down payment and the ability to access credit. This topic is a highly divisive issue in the US political landscape.




Even though home ownership appears to be a shared value across party lines, it is generally purported that Republican positions in terms of regulation and government presence in mortgage markets are much laxer than Democratic positions \citep{hall2022does}. For example, according to recent surveys, two-thirds of both Republicans and Democrats agree that owning a home is necessary to live the American Dream and 73\% of Republicans and Democrats believe that owning a home increases a person's standing in the local community.\footnote{See \url{https://www.prnewswire.com/news-releases/homeownership-is-a-shared-value-across-party-lines-300553639.html}} Republicans typically favor easier access to personal credit and lower banking regulations, at the cost of higher exposure to downside risk and higher individual liability. For example, the republican platform for the 2020 election was pushing toward incentives for accessing mortgages, regulatory downsizes, and minimizing the federal role in zoning decisions; while the Democratic platform promised robust investments in affordable housing production and rental assistance through the development of the Federal Housing Administration (FHA).\footnote{See \url{https://nlihc.org/resource/democratic-party-and-republican-party-platforms-address-affordable-housing}.}



Historically, both parties took clear opposite positions on the regulation of credits and mortgages. For example, the Glass-Steagall Act which separated commercial and investment banking and increased banking regulation in 1933 was written by two Democratic representatives, while the repeal of the Glass-Steagall Act in 1999 by The Gramm-Leach-Bliley Act was written by three Republicans. Similarly, the Dodd-Frank Act of 2010 which tightened financial regulation and increased consumer protection  was written by Democrats and received strong republican opposition. This latest Act was partly dismantled under the Trump administration with the objective to ease mortgage loan data reporting requirements for the overwhelming majority of banks (as it reduced the number of banks subject to heightened regulatory scrutiny). In view of these examples, one can assume that access to mortgages and credit is strongly linked to political behavior as it is a clearly polarized issue over a central topic for most US citizens.

In this paper, we uncover such a link by directly connecting the ability to borrow to voting behavior. Specifically, we assess the effect of the proportion of individuals around salient credit score thresholds on the share of voters for each party at the county-by-congressional-district level. We show that voters tend to favor Republican candidates in districts that contain more voters around specific credit thresholds, where the probability and size of credit differ substantially within few score points, i.e. below such thresholds access to credit tightens substantially. We interpret this result as arising from higher uncertainty in access and quantity of credit that favors Republican candidates who typically run on platforms of deregulation and easier access to credit. The credit uncertainty and its impact on access to home ownership might also favor more conservative candidates through cultural channels as voters tend to be more attracted to conservative rhetoric in times of economic hardship, for example by blaming minorities or foreigners for limiting social mobility (see \cite{FUNKE2016} or \cite{algan-al-2017}), or through its effect on the fear of loss of status (\cite{mutz2018} and \cite{guriev2022}). The recent literature introducing social identity in voting models (\cite{Bonomi-al-21} or \cite{Grossman-21}) shows how adverse economic shocks may generate both a behavioral response that strengthens one’s identification with a specific social group -- e.g. the white working class -- and material interests. Higher economic uncertainty may therefore increase the political relevance of racial and ethnic identities among voters, along with support for culture-based politics, where nationalist and tribal sentiments are salient.




Our empirical approach relies on several data sources. First, we  use  proprietary data on a random 1\% sample of the US population with a valid credit score in 2010, and yearly reports (drawn on June 30th of each year) for the same population from 2004 to 2016 (similar data are described in  \cite{LeevanderKlaauw2010}). This extensive database of about 2 million individuals, observed for 13 years, allows us to construct new data-driven credit score thresholds below which credit tightens substantially. Specifically, we determine the credit score values around which individual total credit amount (including mortgages) jumps the most, for each local labor market (commuting zone) and year over our period of analysis.\footnote{We focus on  thresholds in the credit score range of 560-650 as it is usually perceived to be a  relevant range for uncertain mortgage and credit access, for example, 30.1\% of new mortgages were originated in that range up to 2008, after which that figure dropped to 17.5\% between 2008 and 2016 (\url{https://www.newyorkfed.org/medialibrary/interactives/householdcredit/data/pdf/HHDC_2015Q3}).  This range encompasses subprime and near prime individuals \url{https://www.experian.com/assets/consumer-information/product-sheets/vantagescore-3.pdf}. Prime  individuals are inframarginal with respect to credit availability, as their approval odds are quite high.}
These thresholds are highly relevant as the total amount of credit increases by 20,000USD on average when crossing the threshold by 1 point (where the average total amount of credit is around 100,000USD over the whole sample). We then compute the share of individuals around these thresholds -- we experiment with bandwidths of 5 to 25 credit score points, at 5-point increments -- for each of the 4931 US county-by-congressional districts. We then exploit the quasi-exogenous variation in these shares, across space and time, to assess their effect on US Congressional elections and ideological position.
We use data on US elections at this same level of aggregation from \cite{leip2017david}, focusing on US Congressional candidates to exploit the considerable number of candidates in these races. The information on their ideological positioning on the liberal-conservative spectrum is elaborated through DW-NOMINATE scores \citep{poole1985spatial, poole1991patterns}. A wealth of ancillary datasets complete the analysis: demographic information from the US Census Bureau at the ZIP Code Tabulation Areas (ZCTA) level as well as geographic relationship files to build crosswalks between different geographic units, and exposure to international trade in each local labor market \citep{autor2020importing}.

Our identification strategy exploits the county-by-congressional district's variation in the share of marginal individuals.  Our empirical specifications account for time-invariant local characteristics and state-by-year variation and also control for population traits at the same level of geographical variation  (such as race and gender). We also account for the (instrumented) share of imports from China in each district, as it is one of the major economic drivers for republican votes identified in the recent literature \citep{autor2020importing}. The main identifying assumption is therefore that the variation in the number of people clustered around a specific credit threshold is an exogenous measure of credit access. Our identification argument is firstly based on the objective fact that individuals do not know where these thresholds are and therefore cannot easily manipulate their credit score to be on either side. Banks do not explicitly tell their customers where those thresholds are, and marginal customers would not have full control of their score to the 1-5 points level. If someone has a credit score within 5 points of an unknown threshold it would be very hard to believe that they can exactly infer the threshold and push their score just above it. Our argument is also strengthened by the existing evidence provided by \cite{AgarwalChomsisengphetMahoneyStroebel2018}. Furthermore, our threshold computation design implies that the share of individuals around a specific salient threshold is not generated mechanically by a simple distributional shift of credit score (as thresholds are computed for each year and each commuting zone). This granularity in the data allows us to infer that credit access uncertainty is the driving factor of differential voting behavior and not, e.g., a decrease in all credit scores in the area or an increase in some part of the distribution of credit score that would generate an income effect. We discuss the methods used to infer the thresholds in detail in section \ref{sec:thresholds}.


Leveraging the variation in the share of potential voters around the estimated thresholds we find that a 10pp increase in the share of marginal (potential) voters increases the republican votes by 2.7pp, and reduces that of democrats by 2.6pp. These are sizeable, and robust effects (see Section \ref{sec:voting}), at the margin they would determine winners. Further, we show that the results are not driven by the share of potential voters below vs. above the thresholds, it appears that it is the uncertainty of being around it that matters. Using the DW-Nominate score --  computing the ideology of candidates on social and economic issue based on their roll-call -- we also show a substantial conservative shift for elected candidates in areas with a higher share of voters facing credit-uncertainty (especially democrat candidates). We interpret this result as consistent with the 'cultural channel' as candidates of both parties tend to become more conservative (including on social dimensions) in areas with a higher share of people experiencing credit uncertainty.


Our paper is at the crossing of several literature. First, we contribute to the literature linking economic uncertainty or hardships with a drift toward conservative voting. A large set of papers study the impact of trade (i.e. \cite{colantone-stanig-2018}, \cite{caselli-al-2020} or \cite{autor2020importing}) and economic crisis (\cite{FUNKE2016}, \cite{guiso-al-2017}) in explaining the rise of conservative or populist candidates or populist platforms (\cite{becker-al-2017}). A recent paper (\cite{kara-2022}) also shows that banks reduce the supply of mortgage loans when policy uncertainty increases in their headquarters states.\footnote{Notice that our design is not subject to the potential reverse causality issue implied by this result as we do not exploit the variation of the credit threshold per se.} Finally, \cite{mianal2010} show that representatives whose constituents experience a sharp increase in mortgage defaults are more likely to support the Foreclosure Prevention Act (preventing manipulative foreclosure practices disproportionately harmful to communities of color). Another strand of the literature focuses on linking labor market conditions to conservative voting (\cite{algan-al-2017}) and austerity (\cite{guriev2022}). We contribute to this literature by highlighting the unique role of credit uncertainty on voting behavior. Credit access is a pillar of social mobility and we highlight its role in voting behavior. As it is usually easier to regulate credit markets than manipulate macroeconomic conditions, this has important implications for policymaking.

Second, we speak to the literature that highlights the identity, status, and cultural roots of modern populism. People care deeply about non-monetary factors such as identity, fairness, and status (see \cite{benabou-tirole-2006}; \cite{enke2020}; \cite{guriev2022}, among many others). Some recent literature shows how these sentiments affect voting. \cite{rodrickmukand-2018} highlight the role of "identity politics" which focuses on changing voters' perceptions of who they are. By discussing pride and victimhood, identity politicians create an "in-group" sentiment that helps explain why low-income voters may support a right-wing politician who advocates less redistribution. Along similar lines, \cite{enke2020} provides evidence that the rise of populism is related to the gradual shift of Americans' moral values away from universalist and toward communal ones while \cite{mutz2018} argues that Trump supporters were mostly driven by the threat to their status within the society. \cite{autor2020importing} also support the culture view: the China shock boosts Trump and conservative Republicans' support only in counties with (non-Hispanic) White majorities. \cite{colantone-stanig-2018} produce similar evidence for Europe showing that regions hit hard by Chinese imports are less supportive of democratic institutions and less likely to hold liberal values. We contribute to this literature by showing that credit access uncertainty impacts voting behavior potentially through its effect on social mobility and through the status of home ownership.

Finally, we contribute to the literature identifying the various socioeconomic impacts of credit access: (see \cite{HERKENHOFF2021345} on the impact of consumer credit access on self-employment and entrepreneurship; \cite{NBERw22274} on job finding; and for an extensive discussion on how credit access affects human capital investment and mobility see \cite{HeckmanMosso}).

The rest of the paper is organized as follows. Section \ref{sec:framework} discusses the conceptual framework; Section \ref{sec:data} presents the data used for the analysis. Section \ref{sec:thresholds} details the process of identifying thresholds for credit access; Section \ref{sec:voting} presents the main results; and Section \ref{sec:remarks} concludes.


\section{Conceptual Framework}
\label{sec:framework}

Climbing the ladder of American society  starts with the ability to pursue one's dream: rising through the ranks  as a property owner, small entrepreneur, and the like. The inability to do so because of the lack of capital can push voters towards political parties with a higher propensity to be pro-credit access, pro-business, and with laxer credit regulations. It is generally understood that Republican positions in terms of regulation and government presence are much laxer than Democratic positions. And as such in favor of easier access to personal and business credit. 

We draw a direct connection between credit uncertainty, i.e. the random nature of being approved for credit for marginal individuals, and political choices by the electorate. In particular, we conjecture that voters who are uncertain about their ability to access credit will disproportionately favor republican candidates as they see these candidates' positions as more in line with their needs. Suppose an individual is considering purchasing her first home, a fundamental step towards the dream, and has the need for a mortgage which she would request at the local bank. 
If her credit score puts her in some low and uncertain probability of obtaining such loan, she would support political positions which would increase her approval rates.

Through proximity to discontinuous points in lenders' credit functions, certain individuals face relatively higher uncertainty in obtaining a loan, ceteris paribus. We can disentangle two channels that guide our interpretation of the relationship between credit and voting. When the share of individuals below the salient thresholds of the lenders' credit function has a different effect on voting than the share above (polarization), we can infer that individuals are reacting to being credit constrained. In this case, the true relationship that we uncover is one between the tightness of credit constraints and the demand for higher deregulation. When the share of individuals below and above the salient thresholds have a similar effect on voting, we can instead presume that it is uncertainty in access to credit that relates to demand for deregulation. Our results -- presented in section \ref{sec:voting} -- strongly support the latter.

\begin{figure}[htb!]
	\centering
	\caption{SD of (log) CS at ZIPCODE level (2010)}
		\vspace{-1em}
	\includegraphics[width=1.0\columnwidth]{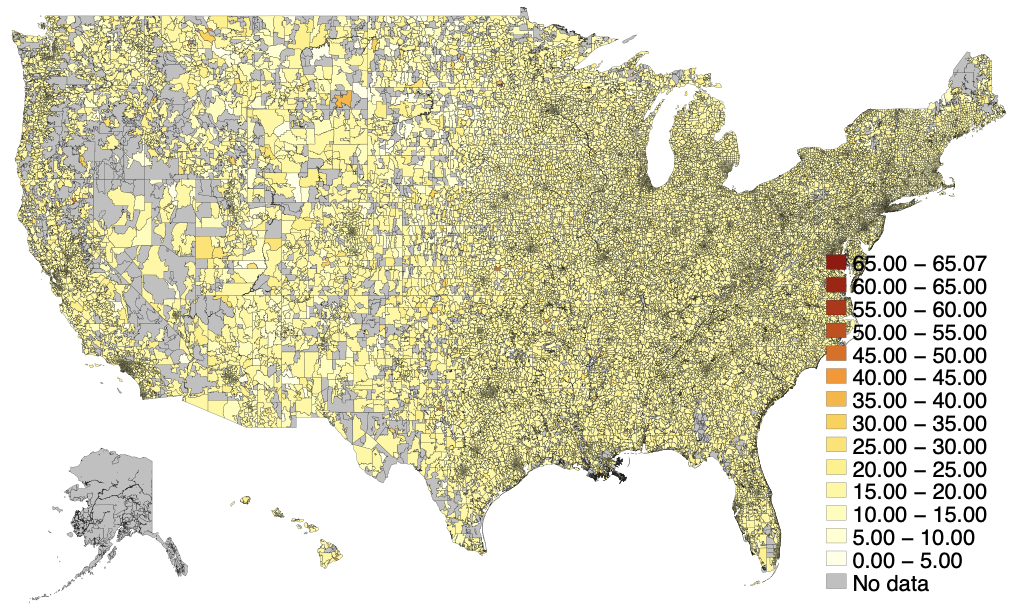}
	\caption*{\footnotesize Notes: Legend is percentage variation within zip code in equally spaced 5pp intervals.}
	\label{fig:sdlncs}
\end{figure}

Our main assumption is that, while the individual is not aware of the exact threshold used by her bank to approve her credit application, she can infer the proximity to the threshold by observing the successes and failures of her peers and her neighbors in obtaining credit. She thus perceives a relatively higher uncertainty in her ability to access credit when close to the salient threshold. 
We rely on the idea that social networks display homophily in characteristics used for computing credit scores and/or spatial correlation in credit scores. In our example, by noting that her social contacts or neighbors had somewhat different success on similar applications to hers (similar credit characteristics and similar homes), she infers that her application will face a substantial degree of uncertainty. This uncertainty would make her more likely to vote for those candidates who are more in favor of laxer requirements and regulations on the mortgage market front in order to increase her chance of obtaining credit. This is a direct economic effect of credit uncertainty on voting behavior. 
In Figure \ref{fig:sdlncs}, we show the standard deviation of (log) credit score at the zip code level in 2010. It appears quite evident that the credit score varies little within zip code (the 50th and 99th percentile are 17\% and  30\% respectively). We find this consistent with our hypothesized \textit{learning-about-uncertain-access} mechanism.\footnote{While variations of at most 30\% are quite limited, one can benchmark that versus the std. dev. of (log) income in the US in 2010 (25-65 years old) from the GRID database project which is 96\% (\url{https://www.grid-database.org/}.}



In view of the recent literature on the political economy of voting (see section \ref{sec:intro}) we also conjecture that credit-access uncertainty -- through its effect on home ownership, self-employment, and social mobility -- might also fuel the fear of status loss. This non-monetary component of individual preference is central to the rhetoric of more conservative candidates in US politics (\cite{guriev2022}) as the fear of status loss can be imputed to minority groups or foreigners following a "us against them" rhetoric. It has been shown that economic uncertainty and financial crises favor such "conservative shift" (\cite{FUNKE2016}, \cite{guiso-al-2017}) and it is likely that the effect is salient for credit-uncertainty given the symbolic role of access to home ownership. Consequently, this "status effect" might also push voters facing credit uncertainty to favor more conservative candidates (democrats and republicans), irrespective of their economic positions on redistribution (as in \cite{rodrickmukand-2018}).


Consequently, we expect that the share of marginal-credit voters will affect the election results: the larger the share of credit-uncertain individuals the larger the voting shares for Republican candidates. We also expect that a larger share of credit-uncertain individuals will translate into more vote shares for socially conservative candidates, republicans and/or democrats.

%


\section{Data}
\label{sec:data}

Our analysis combines several data sources: detailed individual-level data on credit reports from 2004 to 2016 from Experian (section \ref{sec:Data_Experian}); information on electoral outcomes for U.S. Congressional races from \cite{leip2017david} and ideological positions of candidates in the liberal-conservative spectrum \citep{poole1985spatial,poole1991patterns} discussed in section \ref{sec:data_elections}; and several ancillary datasets such as trade exposure in local labor markets \citep{autor2020importing} and Census data (section \ref{sec:data_other}).

\subsection{Experian Credit Reports}\label{sec:Data_Experian}

We have access to proprietary data, provided by Experian, on a random 1\% sample of the US population with valid credit scores in 2010, and yearly reports (drawn on June 30th of each year) for the same population from 2004 to 2016. Similar data are described in detail in \cite{LeevanderKlaauw2010}; and used in \cite{albanesi2018insolvency} for the analysis of the effects of bankruptcy, \cite{albanesi2019predicting} and \cite{DeGiorgiHardingVasconcelos2021} for the prediction of default and death, \cite{HERKENHOFF2021345} to analyze the impact of consumer credit on self-employment and entrepreneurship, in \cite{mian2011house},  \cite{adelino2016credit}, \cite{foote2021cross}, and \cite{albanesi2022credit} to analyze  housing default of 2008. \cite{BachCampaDeGiorgiNosalPietrobon2023} use the same data to study the lifecycle dynamics of credit. The credit report data contain all credit operations of individuals in the formal credit market and include credit scores, number and balances of revolving trades, mortgages, auto loans,  credit limits on the different lines of credit, etc. Further, the data contain  delinquencies and default events on each type of trade (it is common in the industry to refer to credit lines as trades, we will use the two interchangeably). Overall we have over 400 variables describing individual credit behavior for the entire period. In addition, the data contain some basic demographic information, i.e. date of birth, and zip code of residence. 

\begin{figure}[htbp]
	\caption{Shares of individuals across the distribution of credit scores}
	\includegraphics[width=.95\linewidth]{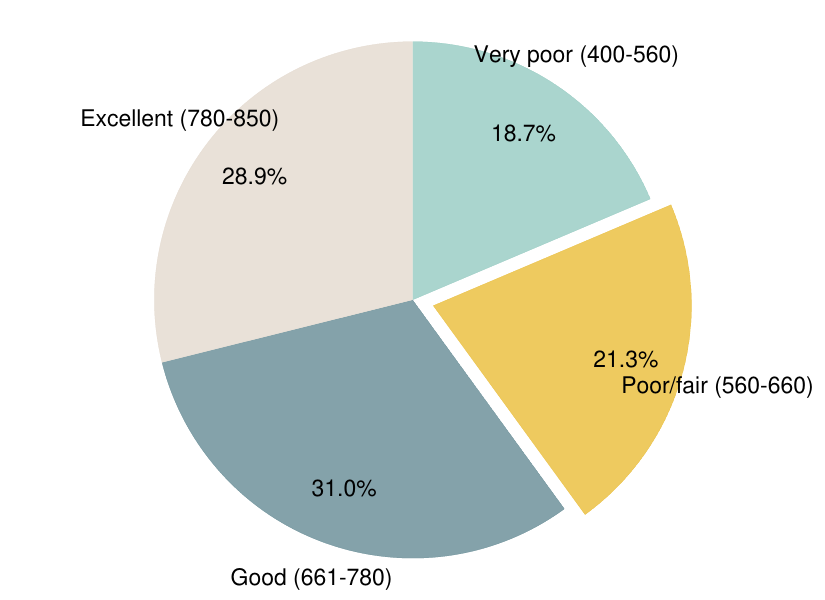}
	\centering
	\label{fig:cs_pie}
	\vspace{-.5cm}
\end{figure}

\begin{figure}[htbp]
	\caption{Average Credit Limits across the distribution of credit scores in the baseline year (2010)}
	\includegraphics[width=.8\linewidth]{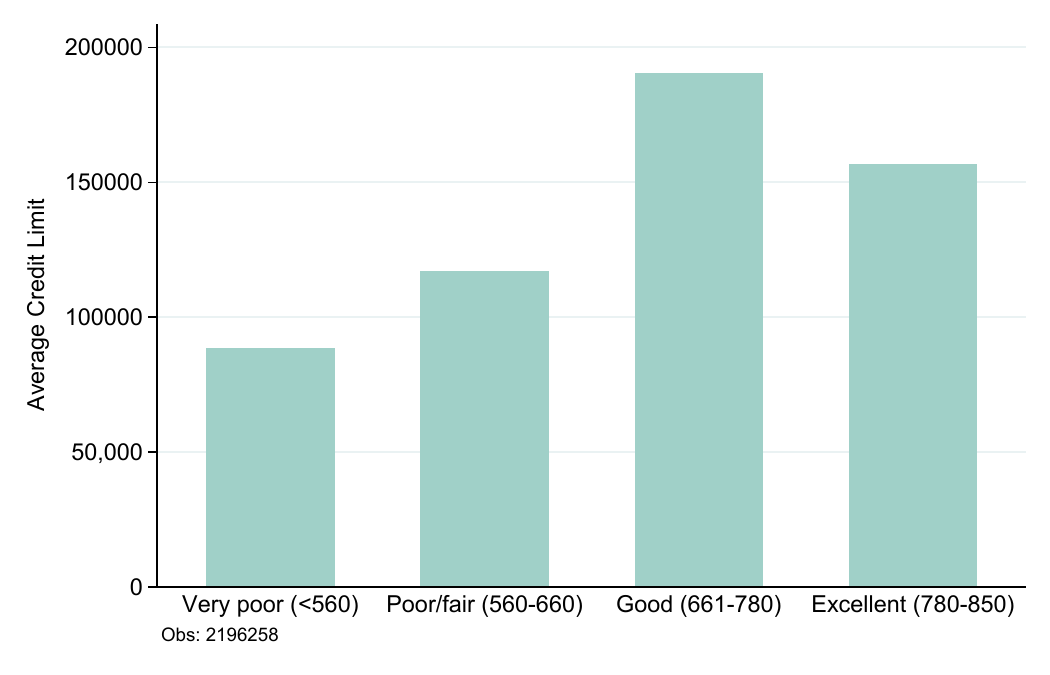}
	\centering
	\label{fig:all5321_descr}
	\vspace{-.5cm}

\end{figure}

These data contain two crucial pieces of information used in the construction of our main explanatory variable, detailed in section \ref{sec:thresholds}: credit scores and total credit amounts. Credit scores are computed using the Vantage Score V3 scoring model and are supposed to predict  the probability of default in the next two years and rank individuals accordingly in a decreasing fashion between 300 and 850 score points. Figure \ref{fig:cs_pie} displays the distribution of credit scores in our data in different categories according to the scoring model. Our focus is on the population of borrowers with poor/fair credit scores as they are more likely to be marginal in the ability to access credit (section \ref{sec:thresholds} elaborates on this choice). The total credit amount on open trades is the total amount of credit on all open accounts that an individual can potentially access (such as revolving credit, mortgages, and other loans).
Figure \ref{fig:all5321_descr} presents the average credit limits for consumers with different credit score bins, while \ref{tab:all5321_descr} provides some additional statistics for the same bins.  Limits differ substantially on average depending on the credit score bin, increasing with credit scores for all groups except for "excellent" borrowers (who are typically less leveraged and older). On average, those with very poor scores have a total limit of about 89,000USD, Poor/Fair about 120,000USD, Good 190,000USD, and Excellent 160,000USD.
While credit scores are continuously distributed in the population, banks discontinuously increase total credit at locally determined thresholds along the credit score distribution (see \cite{AgarwalChomsisengphetMahoneyStroebel2018} and \cite{de2023extension} for such examples).
We focus on such thresholds at the commuting zone and year level as that is the lowest level at which credit markets might vary. Alternatively, we could have identified those thresholds at the State-year level -- the typical level of banking regulation -- however, it is not the appropriate level for the current paper as political candidates run for sub-State seats.\footnote{
An alternative is to explore the probability of successful inquiries for new credit lines, as well as the size of mortgages. In the current paper, we focus on total credit amounts as it is the more complete series, and has the advantage of being determined for the large part on the simple credit score, while for mortgages only the typical approval process uses extensive information on the applicant which are not available to us.}

\begin{table}[htbp]
  \centering
  \caption{Average Credit Limits in USD across the distribution of credit scores in the baseline year (2010)}
  \setlength{\tabcolsep}{10pt}
    \begin{tabular}{lccccc}
    \toprule
    Group (credit score) & Mean  & St. Dev. & Min.  & Max.  & Obs. \\
    \midrule
    Very poor ($<$560) & 88733.98 & 182381.37 & 0     & 15428795 & 230974 \\
    Poor/fair (560-660) & 116946.38 & 201364.94 & 0     & 16246403 & 388954 \\
    Good (661-780) & 190463.22 & 277642.29 & 0     & 17505724 & 619564 \\
    Excellent (780-850) & 156859.70 & 207992.06 & 0     & 16773652 & 647251 \\
    \bottomrule
    \end{tabular}%
  \label{tab:all5321_descr}%
\end{table}%

\subsection{Election Data}\label{sec:data_elections}

Data for electoral outcomes at the county-by-congressional district level between 2004-2016 are from Dave Leip's Atlas of U.S. Presidential Elections \citep{leip2017david}. These data track electoral outcomes over time for Presidential, House, and Senate elections in counties within electoral districts. We have the number of votes obtained by Democrats, Republicans, and other candidates in each electoral year. To maximize the frequency of the data, our main specification focuses on House of Representative elections. We choose these electoral data because the relevant geographic aggregation -- county-by-congressional districts --  allows us to incorporate information on local labor markets in our specification.
We also use Poole-Rosenthal DW-NOMINATE scores to follow shifts in the ideological scores of candidates through roll-call votes in Congress \citep{poole1985spatial,poole1991patterns,mccarty2016polarized}. These scores represent legislator ideologies on a spatial map that allows for comparison across congresses. While several dimensions of these scores are available, we focus on the first dimension which intuitively represents the "liberal" vs. "conservative" divide of American politics. The scores vary from -1 (most liberal) to +1 (most conservative), with values at 0 representing the political center. These measures are widely used in political science and in particular for the study of polarization in the U.S. Congress \citep{persson2002political,bonica2013hasn,bonica2014mapping,fariss2014respect,matsusaka1995fiscal,autor2020importing}.\footnote{For greater detail see: \url{https://voteview.com/about}, \url{https://legacy.voteview.com/page2a.htm}.}
To protect voter secrecy, electoral choices are not available at the individual level. Indeed, electoral outcomes at the precinct level are available (for example, \cite{harvardarchive}), but they do not cover the timeline of our analysis and we were unable to attribute precinct vote shares to zip codes (the finest geographical information in our other data sources) and congressional districts. Other scholars rely on individually elicited voting preferences and behavior, through surveys or registration data. While this information is trackable at the individual level, it usually refers to small samples (N<4000) that are not representative at the county-by-congressional district level or below \citep{enke2020,mutz2018}.


\subsection{Other Data}\label{sec:data_other}

This project leverages a wide array of ancillary datasets that are used for validation and support. To relate the Experian data -- geolocalized at the 5-digit Zip Code Tabulation Area (ZCTA) level -- with the electoral outcomes at the county-by-congressional district level, we build a crosswalk using 2010 ZCTAs to county FIPS relationship files, and ZCTA to Congressional District relationship files, both from the Census Bureau.\footnote{\url{https://www.huduser.gov/portal/datasets/usps_crosswalk.html}} Throughout the paper, time-varying geographic units are restored to 2010 census tracks when possible as it coincides with the relevant census track for the sampling of the Experian data. Several ZCTAs are split across multiple counties. In order to ensure one-to-one mapping with counties, we attribute the zip code to the county where most of its population resides. This approach ensures little distortion as the population is usually concentrated in one county with minor spans over county borders (figure \ref{fig:splitZCTA}). 
We also ensure one-to-one mapping between ZCTAs and congressional districts by looking at official congressional directories for each Congress. We can then easily navigate between counties and congressional districts which do not coincide\footnote{\url{https://www.govinfo.gov/app/collection/cdir}}

We obtain industry exposure to international trade from UN Comtrade, elaborated by David Dorn and described in \cite{autor2020importing} \footnote{\url{https://www.ddorn.net/data.htm\#Industry\%20Trade\%20Exposure}, sections D, E.}
to account for shifts in voting behavior driven by local labor market exposure to international trade in wake of China's entry into the WTO in 2001. These data contain both local labor market exposure to trade with China, as well as trade flows between China and other high-income countries with trade flows comparable to the US.\footnote{The eight other high-income countries have trade data comparable to the US over the full sample period:
Australia, Denmark, Finland, Germany, Japan, New Zealand, Spain, and Switzerland.} These data will be used in the main specification following \cite{autor2020importing} to account for the causal effect of increased exposure to international trade on political polarization. 
Population data from the US Census Bureau  complete the data we use to include information on the racial, gender, and age composition of our geographic units of interest.

A standard concern in studies of political outcomes is gerrymandering, the strategic redistricting of incumbent policymakers after each census to favor the incumbent party in future elections. Following \cite{autor2020importing}, we use  county-by-congressional districts as our main geographic unit, as congressional districts are designed along population criteria and often span across counties and commuting zones. There are ultimately 435 congressional districts that are designed to contain the same number of voters. Section \ref{sec:appendix_gerry} discusses how we break down the data to ensure that the geographic limits of our county-by-congressional district units remain constant.

\section{Empirical design} 

This section presents our main specification and our identification strategy. It also details the construction of our main explanatory variable measuring credit uncertainty and provides insight into the identification strategy.


\subsection{Main specification}

Our aim is to  investigate the relationship between uncertainty in credit access and voting behavior. We conjecture that voters' uncertainty about their ability to access credit, represented by their proximity to ``random'' thresholds that we detail in subsection \ref{sec:thresholds}, pushes voters toward candidates that push for laxer credit market regulation and more populist ones.
 
In order to establish such a link, we estimate the following  equations:

\begin{eqnarray}\label{eq:baseline}
\mathtt{Vote\ Share}_{c,t}&=& \alpha_1+\beta_1 \mathtt{Share\ at\ Thresholds}_{c,t} + \gamma_1 \mathbf{X}_{c,t} + \mathbf{D}_c  + \mathbf{D}_{t} + \epsilon_{1,c,t} \\
\mathtt{Vote\ Share}_{c,t}&=& \alpha_2+\beta^b_2 \mathtt{Share\ Below\  Thresholds}_{c,t} \nonumber \\  && +\beta^a_2 \mathtt{Share\  Above\  Thresholds}_{c,t}+ \gamma_2 \mathbf{X}_{c,t} + \mathbf{D}_c  + \mathbf{D}_{t} + \epsilon_{2,c,t} \label{eq:baseline2}
\end{eqnarray}

\noindent where the left-hand-side variable, $\mathtt{Vote\  Share}_{c,t}$, is the Republican (or Democrat) vote share in county-by-congressional district $c$ and election year $t$. Our analysis covers the US county-by-congressional house elections over the 2004-2016 period. 
$\mathtt{Share\  at\  Thresholds}$, $\mathtt{Share\  below\  Thresholds}$, and $\mathtt{Share\  above\  Thresholds}$ denote the share of voters around, below, and above the most salient credit score thresholds in district $c$ and year $t$ (described in section \ref{sec:thresholds} and Table \ref{tab:descr_mainx}). 
$\mathbf{X}_{c}$ denotes a set of time-varying district-specific covariates, such as the gender and race composition of the district and the (instrumented) share of China imports in the district (described in section \ref{sec:data}). $\mathbf{D}_{c}$ and $\mathbf{D}_{t}$ denotes the full set of county-by-congressional district and time fixed effects, respectively. The choice of this specification is made necessary because we do not observe individual voting behavior and candidates vary within congressional districts, while credit markets likely vary at most at the local labor market level (i.e. groups of adjacent counties). By aggregating vote counts, personal finance information, and demographic data at the county-by-congressional district level, we are effectively estimating equations \eqref{eq:baseline} and \eqref{eq:baseline2} at the finest possible level. 

The $\beta$s are our main coefficients of interest, representing the causal relationship between increases in the share of individuals who are uncertain in their ability to borrow and electoral outcomes. We exploit time and geographic variation in these shares driven by exogenous supply-side shifts in lenders' credit functions to identify such effects. Section \ref{sec:thresholds} describes how we identify and validate the discontinuities in lenders' credit functions, while appendix \ref{sec:appendix} complements the framework with extensive sensitivity analyses of these methods.

An alternative specification uses DW-NOMINATE scores measuring candidate ideological positions along the traditional progressive-conservative spectrum as outcomes of interest.
Except for the left-hand-side variable, everything else remains unchanged. The equations with the ideology measures as outcomes are additionally estimated separately for democrat- and republican-winning districts to detect polarization, i.e. shifts away from the political center. These specifications complement the vote share equations as they elicit the effect of uncertainty in access to credit on the intensity of ideological positions beyond bipartisanship.


\subsection{Credit Thresholds}
\label{sec:thresholds}

A crucial step of our analysis is to find plausible sources of exogenous variation in credit access to then build our main explanatory variable: the share of individuals close to the threshold. We interpret proximity to a salient threshold as generating high uncertainty in credit access. To identify the thresholds for credit access we employ an approach similar to \cite{AgarwalChomsisengphetMahoneyStroebel2018}.
 
\begin{figure}[htb!]
	\centering
	\caption{
		Thresholds: density of individuals and total credit amount}
	\includegraphics[width=1.0\columnwidth]{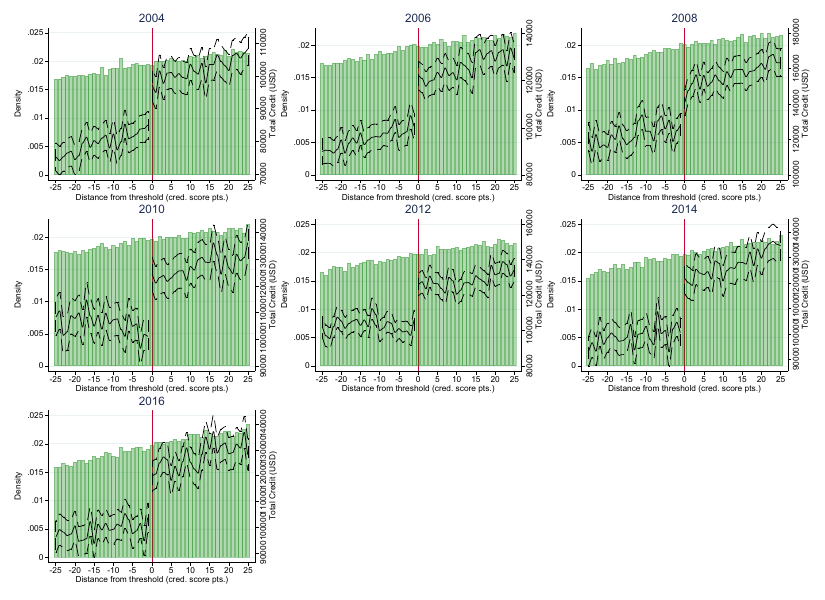}
	\label{fig:rdplot_freq}%
\end{figure}

Variation in access to credit along the values of credit scores is plausibly exogenous as the density of credit scores is smooth at the thresholds (Figure \ref{fig:rdplot_freq} and Figure \ref{fig:balance_newthresh}) and credit scores are locally volatile, such that individuals cannot exactly manipulate their score. The exact formula for the computation of the credit score is proprietary to Vantage. While it is known what type of credit information goes into that calculation, consumers cannot exactly point towards a specific score.\footnote{\url{https://vantagescore.com/press_releases/the-complete-guide-to-your-vantagescore/}} In addition, it is very common to have one's credit score move up or down by 5 points even within a month despite little changes in fundamental behavior, for example, that would happen if there is an increase or fall in credit card balances (even by small amounts), opening or closing credit lines, etc.

The thresholds are identified with a simple regression discontinuity setup at the commuting zone level in each election year.\footnote{Our data would allow us to estimate the thresholds at a yearly frequency. As the outcome of interest in the main specification varies biannually (electoral results), we focus on identifying the thresholds at this level. To ensure sufficient power in sparsely populated commuting zones, we exploit observations from both election and non-election years for estimation.} We test for discontinuities in the total credit limit at five-point intervals in credit scores and we define the thresholds that determine discontinuous access to credit as the credit scores for which we detect an increase in the credit limit in each local regression. Eventually, we compute the explanatory variable used in the main specification as the share of individuals close to the threshold with respect to the total population in each commuting zone and election year.

We run the regressions in each commuting zone $CZ$ and election year $t$. Let $c_i$ be the credit score of individual $i$, $y_i$ be the total credit limit, $d$ be a dummy equal to 1 if it is above the cutoff value of the credit score $\bar{c}$ and 0 otherwise, and $\mathbf{D}_c$ denote county fixed effects. Following \cite{cattaneo2016inference}, we estimate
\begin{equation}\label{eq:thresholds}
    \sinh^{-1}(y_i)=\begin{cases}
    \alpha d_i+\beta_{01}(c_i-\bar{c})+\beta_{02}(c_i-\bar{c})^2+...+\beta_{0p}(c_i-\bar{c})^p+\mathbf{D}_c & \text{ if } c_i<\bar{c} \\
    \alpha d_i+\beta_{11}(c_i-\bar{c})+\beta_{12}(c_i-\bar{c})^2+...+\beta_{1p}(c_i-\bar{c})^p+\mathbf{D}_c & \text{ if } c_i\geq\bar{c} 
    \end{cases}
\end{equation}

for all $\bar{c}$ in 5-point intervals between 560 and 660, election year, and commuting zone.\footnote{Cutoffs $\bar{c}$ vary at the commuting zone and yearly level ($(CZ,t)$ in our notation). For simplicity, we drop these subscripts hereafter.} The degree of the polynomial transformation $p$ is optimally chosen to allow for the potential outcome to have some direct dependence on the credit score and is equal to 4 in our setting \citep{CalonicoCattaneoTitiunik2014,CattaneoJanssonMa2020}. We interpret positive and statistically significant $\alpha(\bar{c})$ as discontinuous and locally exogenous increases in access to credit where $\bar{c}^*$ is the threshold.
Commuting zones represent local labor markets within which banks decide their credit functions competitively and are appropriate markets due to the large number of individuals who access banking services through credit unions which are local in nature.\footnote{Between 2004 and 2016, 94 million customers on average were members of credit unions. Even though the number of credit unions has been decreasing, there were 7944 CUs on average \citep{NCUA2016}.} Furthermore, as credit-constrained individuals are mostly located at the bottom of the distribution of credit scores, they are more likely to be serviced by CUs that guarantee minimum financial services \citep{creditinvisible}. Although banks need not choose discontinuous credit functions, the data and the literature suggest this is the case \citep{AgarwalChomsisengphetMahoneyStroebel2018,de2023extension}. 
To ensure sufficient power, we restrict our analysis to commuting zones with more than 500 observations. We further restrict the portion of the distribution of credit scores within which we search for thresholds in the 560-660 interval. We choose 660 as the upper limit of our interval of interest because it represents the cutoff between "fair" and "good" credit scores in the Vantage credit score model.\footnote{\url{https://www.experian.com/blogs/ask-experian/credit-education/score-basics/what-is-a-good-credit-score/}} We do not seek thresholds above 660 because fewer individuals would be marginal at these values. We do not seek thresholds below 560 because the distribution of credit limits is distorted by the fact that individuals who face solvency issues are pushed down to credit scores around 525 independently of their previous credit score, which creates bunching and negative thresholds because of extraordinary behavior, such as declaring bankruptcy. 21.3\% of individuals in our dataset exhibit credit scores in this range (as a reminder figure \ref{fig:cs_pie} provides additional information on the distribution of credit scores).

We successfully run 59,237 regressions, i.e. in election year-commuting zone units with more than 500 observations. For each election year, we have between 395 and 410 commuting zones and 298-316 valid thresholds $\bar{c}^*$, i.e. cutoffs for which $\alpha_{\bar{c}^*}$ is positive and statistically significant. In certain instances, we identify positive and statistically significant thresholds in certain years and not in others. When a commuting zone presents discontinuities in access to credit in one year, we impute the missing election years using previous observations under the assumption that banks' decisions to grant credit according to discontinuous credit functions do not vary between election years. In a few instances, we are never able to detect discontinuities over time for certain commuting zones and thus disregard them. Table \ref{tab:CZ_thresh} summarizes these results and the available thresholds. 
We also detect multiple or contiguous thresholds in certain commuting zones and election years. In these cases, we only keep the threshold with the largest coefficient, thus implying the largest jump in credit limit. We identify thresholds according to this conservative rule to reduce the chance of detecting "false thresholds" due to discontinuity detection bias, for example, due to variation in the density of borrowers around the thresholds and discontinuities in credit demand. 

\begin{table}[htbp]
    \setlength{\tabcolsep}{5pt}
  \centering
  \caption{ \centering Commuting Zones (CZ) and valid thresholds}
    \begin{tabular}{cccc}
    \toprule
    \multicolumn{1}{l}{Election Year} & \multicolumn{1}{l}{Total CZ} & \multicolumn{1}{l}{CZ with thresh.} & \multicolumn{1}{l}{CZ with thresh. (imp.)} \\
    & (1) & (2) & (3) \\
    \midrule
    2004  & 409   & 302   & 408 \\
    2006  & 410   & 303   & 408 \\
    2008  & 408   & 312   & 408 \\
    2010  & 403   & 309   & 403 \\
    2012  & 396   & 311   & 396 \\
    2014  & 400   & 316   & 400 \\
    2016  & 395   & 298   & 395 \\
    \bottomrule
    \end{tabular}%
  \label{tab:CZ_thresh}%
  \caption*{\footnotesize (1) -- total CZs with sufficient observations (i.e. $>500$), (2) -- number of CZs for which we detect a positive and statistically significant $\bar{c}^*$, (3) -- number of CZs for which we are able to impute missing thresholds by keeping previous ones in years in which no valid threshold was detected.}
\end{table}%

\begin{figure}[htb!]
	\centering
	\caption{
		\vspace{-1em}
		Frequency of thresholds $\bar{c}^*$ }
	\includegraphics[width=.8\columnwidth]{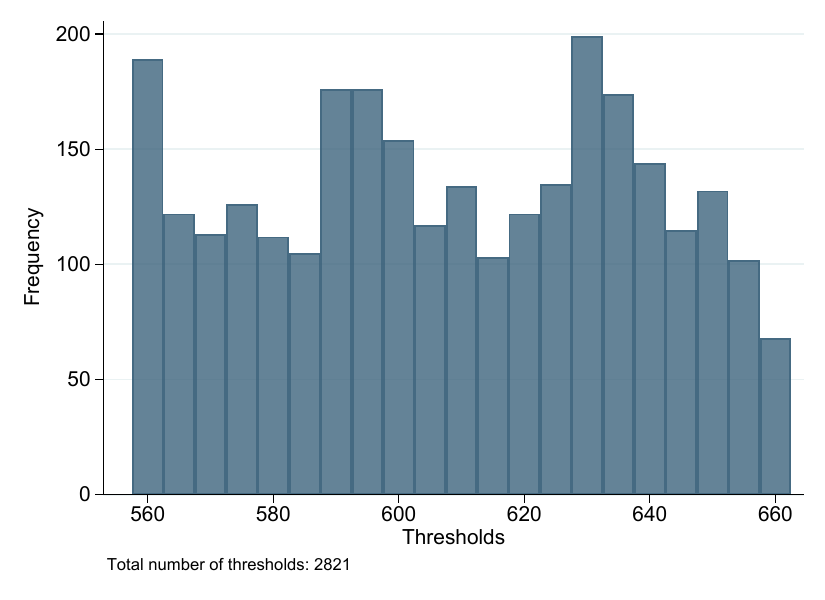}
	\label{fig:hist}%
\end{figure}

Figure \ref{fig:hist} provides additional information on the location of the 2,821 thresholds $\bar{c}^*$. 
While thresholds around 560 and 630 are more frequently observed in the dataset, all values of $\bar{c}^*$ are present. 

Figure \ref{fig:rdplot_freq} displays the (average) magnitude of the discontinuities in the total credit limit that we detect around the thresholds using equation \eqref{eq:thresholds}. For each election year, we center observations around the relevant threshold in the individual's commuting zone and plot the density of individuals in each credit score point (green bars, left axis). While the density slightly increases with higher credit scores, there is no bunching of individuals just above or below the thresholds, so the density is smooth at the threshold (\cite{mccrary2008manipulation} test). The black lines reflect our estimates of credit limits from equation \eqref{eq:thresholds} with 95\% CIs, transformed back into dollar values (right axis). In all election years, the discontinuity in credit limits around the thresholds is statistically significant and large, increasing on average by 20,000USD around the thresholds. On average, valid thresholds have a coefficient $\alpha=1.14$ (inverse hyperbolic sine scale). This implies an increase in the credit limit of 23,400USD as low as the 25th percentile of the distribution of borrowers (from 11,013USD to 34,435USD), and larger increases thereafter. As credit limits are automatically updated by lenders and the jumps are economically large, it is unlikely that discontinuity detection biases alone explain these jumps. Figures \ref{fig:rdplot} and \ref{fig:balance_newthresh} in the appendix provide additional evidence on the discontinuity of credit limits (Figure \ref{fig:rdplot}) and of the smoothness of credit scores (Figure \ref{fig:balance_newthresh}) around the centered thresholds.

\paragraph{The main explanatory variable.} Once we have identified the thresholds $\bar{c}^*$, we can build the main explanatory variable as the share of individuals with a credit score close to the threshold with respect to the total population in the county by congressional district cell as used in equations \eqref{eq:baseline} and \eqref{eq:baseline2}. No crosswalks are needed as commuting zones are groups of contiguous counties that form local labor markets and thus fully contain our geography of interest, county-by-congressional districts (CCD). The main assumption that underlines our identification of the discontinuities in credit limits is that being above or below the threshold is locally random. In fact, it is actually weaker than that, we simply need that the share of people around those thresholds are quasi-random. That appears rather plausible as consumers do not know where the thresholds are. 

As we have no specific size for selecting the neighborhood of the thresholds in terms of credit score points, we compute the shares of individuals close to the thresholds within 5, 10, 15, 20, and 25 credit point deviations above, below, and in total. In our preferred specification, we will analyze shares with a bandwidth of 15 credit score points. Table \ref{tab:descr_mainx} summarized the population shares of individuals above, below, or close  to the thresholds. Not surprisingly, the shares increase as the bandwidth of credit score points considered increases. As expected, roughly the same amount of individuals are above and below the thresholds, further supporting that there is no bunching. In the preferred specification, 6.6\% (std.dev. 4.2\%) of individuals are within 15 credit score points of the threshold. 

\begin{table}[htbp]
  \centering
  \caption{Population shares close to 500s thresholds at different bandwidths (BW) and weighted by population (14,549,479 total observations)}
  \setlength{\tabcolsep}{12pt}
    \begin{tabular}{lcccc}
    \toprule
    Variable & Mean  & St. Dev. & Min   & Max \\
    \midrule
    share(tot), BW: 5 & 0.023 & 0.026 & 0     & 1 \\
    share(above), BW: 5 & 0.013 & 0.020 & 0     & 1 \\
    share(below), BW: 5 & 0.011 & 0.016 & 0     & 1 \\
    share(tot), BW: 10 & 0.045 & 0.035 & 0     & 1 \\
    share(above), BW: 10 & 0.024 & 0.026 & 0     & 1 \\
    share(below), BW: 10 & 0.021 & 0.022 & 0     & 1 \\
    \textbf{share(tot), BW: 15} & 0.066 & 0.042 & 0     & 1 \\
    \textbf{share(above), BW: 15} & 0.035 & 0.030 & 0     & 1 \\
    \textbf{share(below), BW: 15} & 0.032 & 0.028 & 0     & 1 \\
    share(tot), BW: 20 & 0.088 & 0.048 & 0     & 1 \\
    share(above), BW: 20 & 0.046 & 0.034 & 0     & 1 \\
    share(below), BW: 20 & 0.042 & 0.032 & 0     & 1 \\
    share(tot), BW: 25 & 0.110 & 0.054 & 0     & 1 \\
    share(above), BW: 25 & 0.058 & 0.037 & 0     & 1 \\
    share(below), BW: 25 & 0.052 & 0.036 & 0     & 1 \\
    \bottomrule
    \end{tabular}%
  \label{tab:descr_mainx}%
\end{table}%

\subsection{Discussion on identification.}\label{sec:identification} Section \ref{sec:thresholds} clarifies how we identify the cutoff points, the points in the credit score distribution -- for poor and fair borrowers -- where a few score points make a substantial difference in the probability of getting a loan and its size. Over the whole sample, we estimate an average 20,000USD jump in the total credit available to individuals on either side of the cutoffs. These jumps are substantial, typically larger than 20\% of the total credit limit. Customers/voters do not know where these thresholds are exactly, they are typically not communicated and depend on the internal model of banking used by each bank at the local level. Banks also experiment with such thresholds in order to acquire new customers (\cite{de2023extension}). 
What is relevant for our analysis is that voters in those proximities are aware that obtaining a loan is not a done deal. In fact, the outcome of a mortgage or credit application is rather uncertain as it will depend on where exactly the thresholds are. Exposed customers face substantial uncertainty in their ability to borrow. We hypothesize that while customers do not exactly know where the cutoffs are, they are aware that they are in a region of the credit score distribution where obtaining additional credit is in doubt, and therefore operate under larger uncertainty than consumers with larger, inframarginal, credit scores. Importantly, the fact that the discontinuities are not known by the individuals does not pose a threat to the identification of the thresholds, as the lack of common knowledge about the thresholds' location does not affect the efficiency of the treatment effect estimator asymptotically \citep{porter2015regression}. 

We remind the reader that shifts of a few score points within a short period of time are rather frequent, such that the monthly individual variation for those between 560 and 660 credit score points is typically plus/minus 5 points without much action on the credit side. While customers have some control over their scores, they cannot pinpoint them. In addition, they have no knowledge of where exactly the thresholds are and when exactly their credit score will change, as it depends on when the relevant financial institutions send in their reports to the credit bureau. 
Furthermore, we note that credit score apps, and costless credit score verification, were not overly present before 2010, and checking personal scores would have been costly as a small fee would be charged. As itemized credit reports were not widely available, actions to improve credit scores were also harder to single out.\footnote{For a brief history of the leading app in this market, Credit Karma, see \url{https://www.creditkarma.com/ourstory}}. 
Consequently, credit-constrained individuals who face uncertainty in their ability to borrow would inquire frequently with the lenders and thus observe their own success rate and that of their peers (although imperfectly).
Because of the frequent variation of the credit score with respect to the threshold and the infrequent nature of elections, we believe that the relative position of the individual with respect to the threshold -- as observed in June of the election year -- is effectively random.

Finally, the $\beta$s in equations \eqref{eq:baseline} and \eqref{eq:baseline2} are our coefficients of interest. They inform us of how uncertainty in access to credit drives shifts in the vote margin and the ideological position of winning candidates through variation across time and space in the share of individuals around the thresholds. We allow for the effect of uncertainty on polarization to be heterogeneous depending on the share of individuals below and above the threshold, separately (equation \eqref{eq:baseline2}). The separation of specifications \eqref{eq:baseline} and \eqref{eq:baseline2} guides the interpretation of the coefficients. 
Heterogeneous behavior above and below the threshold ($\beta^b_2$ and $\beta^a_2$) is consistent with individuals knowing their position relative to the threshold, therefore eliciting the effect of exogenous variation in credit constraints on voting. Homogeneous behavior is consistent instead with uncertainty in credit access: the small fluctuations in credit scores that affect the individual's borrowing capacity cannot be controlled and individuals notice that similar borrowers obtain higher/lower credit limits and cannot control their status relative to them. Our results showcase homogeneous above- and below-threshold behavior and are thus consistent with the uncertainty in access to credit narrative. These findings are discussed in-depth in section \ref{sec:voting}.



\section{Voting}
\label{sec:voting}


\subsection{Main results - Voting Shares} 

Table \ref{tab:main} displays our baseline results produced by the estimation of equation (\ref{eq:baseline}). Population weights with the counts of individuals in each cell used to compute the shares of individuals close to the thresholds are included and all standard errors are clustered at the county-by-congressional district level. Column 1 displays the effect on Republican vote share (in percentage) of a variation in the share of people just above (line 2) and below the credit threshold (line 3) by a margin of 15 credit points. Column 2 displays the effects on republican voting of an increase in the share of people within a margin of 15 points above and below the credit threshold. Both specifications show a strong and significant positive effect. In county-by-congressional district cells with a higher share of people clustered around the salient credit threshold (with higher credit uncertainty), we observe a higher share of republican votes than in other areas. In particular, a 10 percentage point increase in the share close to the thresholds will increase republican vote share by 2.7pp and decrease democratic votes by 2.6pp. These findings are consistent with the hypothesis that proximity to salient thresholds increases uncertainty in access to credit, as opposed to differentiating between individuals who face varying credit constraints. All columns control for the (instrumented) share of China imports following the specification of local trade exposure in commuting zones derived by \cite{autor2014trade,autor2020importing,acemoglu2016import}, as well as the percentage of white people as well as the share of women in the district.\footnote{We account for shocks in local labor markets driven by the growth of Chinese import penetration by measuring trade flows between the U.S. and China at the local labor market level, accounting for industry structure within each LLM at the beginning of the analysis. To disentangle demand and supply shocks to LLMs, we instrument the import exposure variable with the composition and growth of Chinese imports in eight other developed countries with comparable trade data: Australia, Denmark, Finland, Germany, Japan, New Zealand, Spain, and Switzerland, following \cite{autor2020importing}.} The coefficients for these demographic controls are of the expected signs: a higher share of white and male voters in an area translates into a higher vote share for republicans. The effect is the opposite for democrats, benefiting more from a higher share of non-white and women voters. Interestingly, the main effect of proximity to the credit threshold in column 2 is very similar above and below the threshold in column 1: what seems to matter for voting behavior is the overall proximity to the threshold -- and the uncertainty it generates -- more than the direct economic effect created by the jump above the threshold. 

Columns 3 and 4 display similar specifications for democratic voting. The effects are the close opposite of what we observed in columns 1 and 2, which is not surprising given the almost perfect two-party system save for independent candidates: an increase in the share of people below or above the threshold (column 3) and around the threshold (column 4) is associated with a decrease in democrat vote shares in these areas. As for  column 2, we notice that the effect is similar for the share of people below and above the threshold: the effect is driven by the overall proximity to the threshold and not the crossing of the threshold itself. The coefficient displayed in column 4 implies that an increase of 10pp in the share of people with credit scores located around the threshold implies a decrease in democrat vote shares of 2.6pp.

\begin{table}[htbp]
  \centering
  \caption{Vote Shares and Credit Access Uncertainty}
  \setlength{\tabcolsep}{8pt}
    \begin{tabular}{lcccc}
    \toprule
          & (1)   & (2)   & (3)   & (4) \\
    VARIABLES & Rep   & Rep   & Dem   & Dem \\
    \midrule
          &       &       &       &  \\
    Share close thresh. &       & 0.268** &       & -0.264** \\
          &       & (0.119) &       & (0.118) \\
    Share close thresh. Above & 0.248** &       & -0.258** &  \\
          & (0.126) &       & (0.125) &  \\
    Share close thresh. Below & 0.292* &       & -0.272* &  \\
          & (0.158) &       & (0.156) &  \\
    Share China Import & -0.019 & -0.019 & 0.017 & 0.017 \\
          & (0.014) & (0.014) & (0.014) & (0.014) \\
    Share White & 0.705*** & 0.705*** & -0.698*** & -0.698*** \\
          & (0.020) & (0.020) & (0.020) & (0.020) \\
    Share Female (voting age) & -0.969*** & -0.969*** & 0.967*** & 0.967*** \\
          & (0.183) & (0.183) & (0.179) & (0.179) \\
        Fixed effects & \multicolumn{4}{c}{County x congressional district, year}\\
          &       &       &       &  \\
    Observations & 12,053,931 & 12,053,931 & 12,053,931 & 12,053,931 \\
    R-squared & 0.435 & 0.435 & 0.420 & 0.420 \\
    Mean of dep. var. & 0.464 & 0.464 & 0.505 & 0.505 \\

    \midrule

    \multicolumn{5}{l}{*** p$<$0.01, ** p$<$0.05, * p$<$0.1} \\
    \multicolumn{5}{l}{Cluster robust SE at county-by-congressional district in parenthesis.}  \\
    \end{tabular}%
  \label{tab:main}%
\end{table}%


\subsection{Conservative shift of elected candidates.} Table \ref{tab:nominate} displays results where we look at the effect of the share of potential voters around the credit threshold on the political position of the elected candidate (using the well-known DW-NOMINATE index developed by Keith T. Poole and Howard Rosenthal \citep{poole1985spatial,poole1991patterns,mccarty2016polarized}). The main objective is to test the "ideological channel" purporting that credit uncertainty pushes voters toward candidates using a more conservative rhetoric through, for example, a fear of status loss. Recall that the main dimension of DW-NOMINATE scores locates candidates on a "liberal-conservative" space, ranging from -1 (most liberal) to +1 (most conservative) (see section \ref{sec:data}). To perform this analysis we use the specification in Equations \ref{eq:baseline} where we replace the left-hand-side variable with the DW-NOMINATE index. Columns 1 and 2 present the results for the Republican-winning congressional districts splitting coefficient for the share of citizens above and below the threshold (column 1) and around the threshold (column 2). We observe that the share of people affected by credit uncertainty does not impact the ideology of elected candidates in republican-winning districts, while trade exposure to the "China shock" seems to matter significantly (as in \cite{autor2020importing}). However, columns 3 and 4 show that for Democratic-winning congressional districts, our measure of credit uncertainty explains an increase toward more conservative ideology by a substantial margin. This effect on the ideology of elected candidates is again symmetric with respect to the share of people above and below the threshold. Finally, columns 5 and 6 confirm that the conservative shift of elected candidates triggered by the share of people facing credit uncertainty is true for all congressional districts, irrespective of party affiliation. Column 6 implies that an increase of 10pp of the share of people located around a salient credit threshold increases by 5pp the DW-NOMINATE score of the average elected candidate. This is a substantial increase as it represents a doubling of the mean of the index (equal to 0.051 for all candidates) on the direction of more conservative ideology.


This set of results paints an overall consistent picture: an increase in the share of people facing credit uncertainty is associated with a shift toward the election of more conservative candidates, and this effect is particularly true for Democratic candidates. This substantiates the idea that economic uncertainty fuels conservative rhetoric by reinforcing the "within-group" narratives and identifying external groups as the main causes of economic hardship (e.g. by pointing to the fact that a specific minority group benefits more from a specific housing program). We read this result as consistent with the literature discussed in the Introduction (\cite{Bonomi-al-21},\cite{Grossman-21}) showing that economic shocks have the potential to magnify the political salience of racial and ethnic identity, yielding
an increase in conservative voting on social issues even conditional on
economic status.


\begin{table}[htbp!]
  \centering
  \caption{DW-NOMINATE scores along the progressive (-) -- conservative (+) dimension and exposure to credit uncertainty}
    \begin{tabular}{lcccccc}
    \toprule
          & (1)   & (2)   & (3)   & (4)   & (5)   & (6) \\
    VARIABLES          & Rep. & Rep. & Dem. & Dem. & All & All \\
    \midrule
          &       &       &       &       &       &  \\
    Share close thresh. &       & -0.035 &       & 0.471*** &       & 0.509** \\
          &       & (0.080) &       & (0.145) &       & (0.258) \\
    Share close thresh. Above & -0.134 &       & 0.478*** &       & 0.434 &  \\
          & (0.082) &       & (0.168) &       & (0.269) &  \\
    Share close thresh. Below & 0.104 &       & 0.465** &       & 0.599* &  \\
          & (0.117) &       & (0.186) &       & (0.344) &  \\
    Share China Import & 0.023*** & 0.023*** & 0.014 & 0.014 & 0.030 & 0.030 \\
          & (0.008) & (0.008) & (0.018) & (0.018) & (0.029) & (0.029) \\
    Share White & 0.048* & 0.047* & 0.286*** & 0.286*** & 1.103*** & 1.103*** \\
          & (0.025) & (0.025) & (0.021) & (0.021) & (0.048) & (0.048) \\
    Share Female (voting age) & 0.146 & 0.146 & -0.409* & -0.409* & -1.037** & -1.037** \\
          & (0.184) & (0.184) & (0.237) & (0.237) & (0.434) & (0.434) \\
        Fixed effects & \multicolumn{6}{c}{County x congressional district, year}\\
          &       &       &       &       &       &  \\
    Observations & 6,275,158 & 6,275,158 & 5,867,112 & 5,867,112 & 12,146,679 & 12,146,679 \\
    R-squared & 0.053 & 0.052 & 0.228 & 0.228 & 0.296 & 0.296 \\
    Mean of dep. var.  & 0.454 & 0.454 & -0.380 & -0.380 & 0.051 & 0.051 \\
    \midrule
    \multicolumn{7}{l}{*** $p<0.01$, ** $p<0.05$, * $p<0.1$} \\
    \multicolumn{7}{l}{Clustered SE at county x congressional district, 15 credit score point bandwidth.} \\
    \end{tabular}%
  \label{tab:nominate}%
  \caption*{(1)-(2) ideological measures for Republican elected representatives, (3)-(4) Democrat elected representatives, (5)-(6) all elected representatives. 
  }
\end{table}%

\paragraph{Sensitivity analysis}

We perform a series of robustness exercises on our baseline results in  \ref{tab:App_Dem_bw_abovebelow}-\ref{tab:app_GerryM_nom1}. Tables \ref{tab:App_Dem_bw_abovebelow}-\ref{tab:App_Rep_bw_abovebelow} display results for the share of people above and below the threshold where we vary the bandwidths around the thresholds from 5 to 25 credits points by increments of 5 points (recall that our baseline results are based on a bandwidth of 15 points). Table \ref{tab:App_Dem_bw_abovebelow} displays the results for democratic vote shares and Table \ref{tab:App_Rep_bw_abovebelow} for Republican vote shares. In both tables, the main coefficients remain very stable as we vary the size of the bandwidth. Tables \ref{tab:App_Dem_bw}-\ref{tab:App_Rep_bw} provide  similar robustness checks for the share of people around (above and below) the threshold, for republican and democratic vote shares. We observe in these tables a decrease in the magnitude of the effect as we increase the bandwidth. This is intuitive, as increasing the bandwidth dilutes the uncertainty associated with proximity to the credit threshold and most likely decreases the strength of our main effect.

Tables \ref{tab:App_Nom1_AboveBelow_DR}, \ref{tab:App_Nom1_allshare_DR}, \ref{tab:app_nom1_allcand_allshare} propose  similar robustness exercises, with varying bandwidth, focusing on the DW-NOMINATE score of the winning candidate.  Tables \ref{tab:App_Nom1_AboveBelow_DR} and \ref{tab:App_Nom1_allshare_DR} display results for various bandwidths above/below the thresholds and around the thresholds, for republican and democratic-winning candidates. The results in Table \ref{tab:App_Nom1_AboveBelow_DR} are consistent with our baseline results: increasing the share of individuals around salient credit thresholds tends to be associated with more conservative democrat-winning candidates (above and below). Table \ref{tab:App_Nom1_allshare_DR} shows that the share of people around the threshold (above and below) tends to render republican-winning candidates more conservative. Interestingly, table \ref{tab:app_nom1_allcand_allshare} shows that our main variable of interest has a strong impact on the conservative rhetoric of all elected candidates, republican and democrat. Similarly to the results on vote shares, the effect is smaller as the bandwidth increases.

We then turn toward to possible confounding role of gerrymandering, as some of the congressional districts were redesigned over our period of analysis.  Table \ref{sec:appendix_gerry} shows results only for 2012, 2014 and 2016 elections, where no redistricting took place. Our baseline results are qualitatively robust to this specification, indicating that gerrymandering per se does not drive the results of our main specification. Focusing on those election years for the DW-NOMINATE score (table \ref{tab:app_GerryM_nom1}), all columns show that the main coefficients tend to display a lower significance, while similar in magnitude.


\section{Concluding Remarks}
\label{sec:remarks}

We show a novel role of credit access in determining political and voting behavior. Credit access has a strong economic and symbolic value, as it is an essential component of the American dream through the social ladder of housing, and self-employment. We highlight that individuals with uncertain access to credit -- located around salient credit score thresholds -- appear to disproportionately vote for republican candidates and more conservative candidates overall. 

In particular, we show that a 10pp increase in the share of individuals at the margin of credit access causally increases the vote shares for Republicans by 2.7pp (and decreases democratic vote shares by 2.6pp) a margin which would in several instances change the election results. Further, we show that even in democratic Congressional districts an increase in the share of marginal credit individuals renders democratic candidates' rhetoric  more conservative. An increase of 10pp in the share of individuals around the threshold doubles the average of the NOMINATE score across all candidates, indicating a strong move toward a more conservative ideology. 

We interpret our results as consistent with 2 channels. First, a direct economic channel, as higher uncertainty in credit access might favor Republican candidates who typically run on platforms of deregulation and easier access to credit. Second, our results also support a complementary "cultural channel", as economic uncertainty can amplify the political salience of racial and ethnic identity and create a conservative shift (see \cite{Bonomi-al-21} or \cite{Grossman-21}). This is what we observe with the DW-NOMINATE score, especially for elected Democratic candidates. 

These results bear important policy implications as credit access is usually easier to manipulate at the local level than other macroeconomic indicators (e.g. trade access, GDP, etc.).  Our results also highlight that political conservatism might find part of its origin in credit uncertainty. Legislators can therefore ease access to credit (or reduce credit uncertainty) through local regulations and avoid an unwanted rise in social conservatism and racial tensions.

Future work can look at the exact functioning of the cultural channel. In particular, future research might study in more detail if the DW-NOMINATE score varies differentially with credit uncertainty in areas with different racial and gender compositions. It might investigate if the effect on conservatism is more pronounced in areas with stronger regulation that favors discriminated minorities (e.g. housing regulation).



\bibliography{bib}

\newpage

\appendix

\section{Appendix}\label{sec:appendix}

\setcounter{table}{0}
\renewcommand{\thetable}{A\arabic{table}}

\setcounter{figure}{0}
\renewcommand{\thefigure}{A\arabic{figure}}


\begin{figure}[htbp]
	\caption{Population shares in split ZCTAs}
	\includegraphics[width=.8\linewidth]{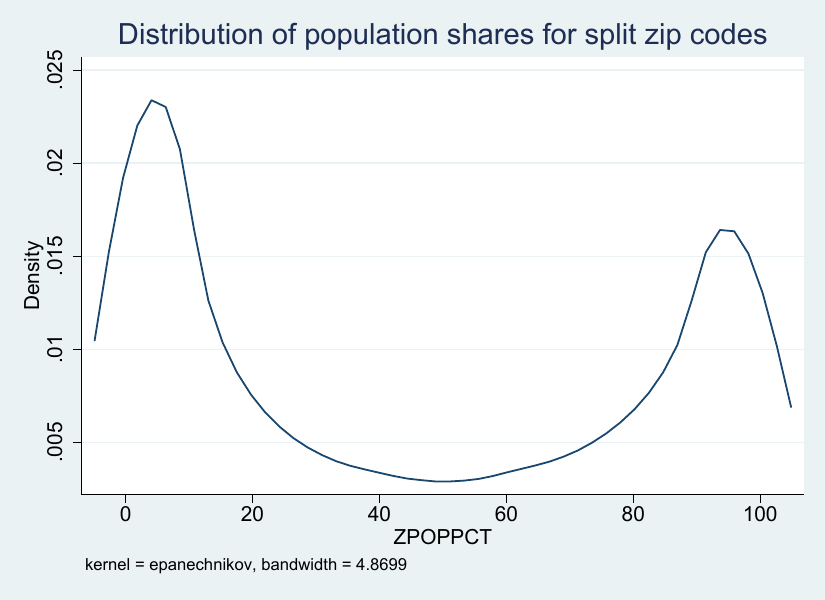}
	\centering
	\label{fig:splitZCTA}
	\vspace{-.5cm}
\end{figure}

\begin{figure}[htb!]
	\centering
	\caption{
		Thresholds: discontinuity in total credit amount}
	\includegraphics[width=1.0\columnwidth]{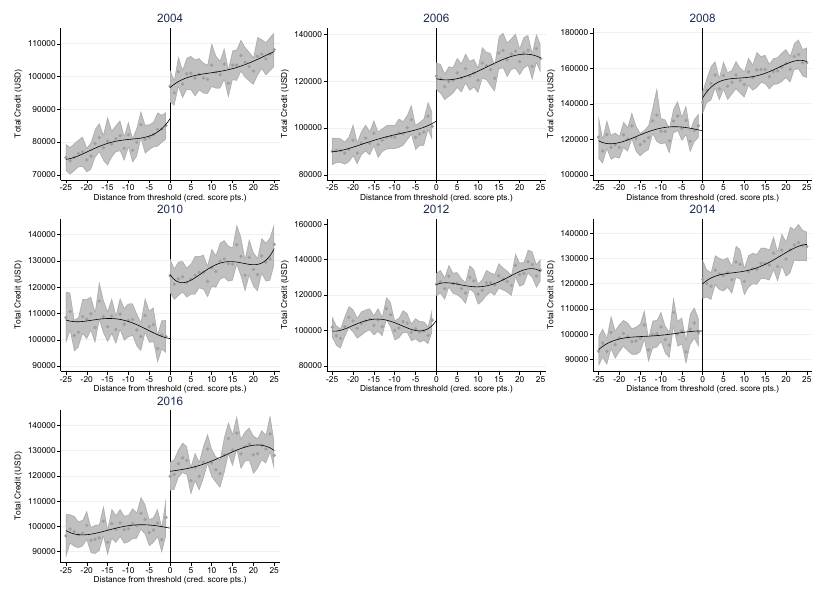}
	\label{fig:rdplot}%
\end{figure}

\begin{figure}[htb!]
	\centering
	\caption{
		Credit score density smoothness around the thresholds over time}
	\includegraphics[width=1.0\columnwidth]{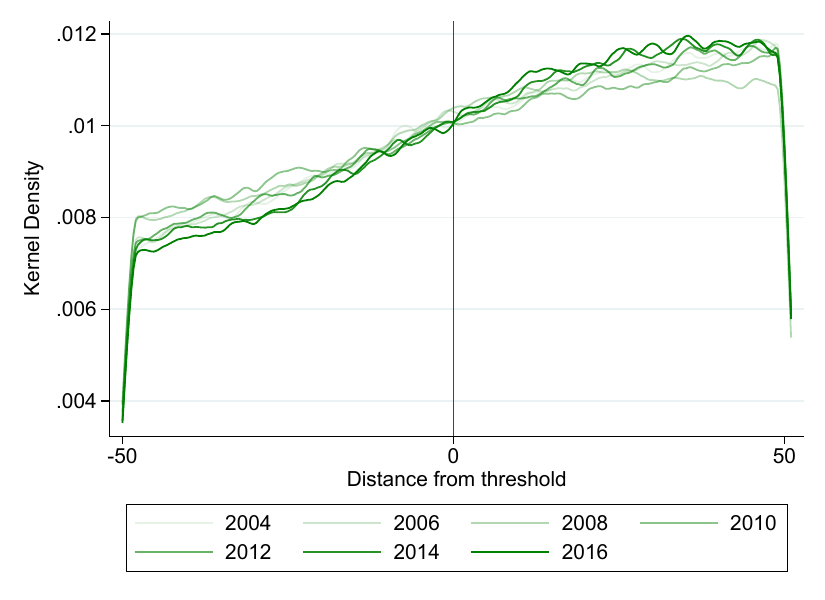}
	\label{fig:balance_newthresh}%
\end{figure}    

\FloatBarrier

\subsection{Additional Results}

\begin{table}[htbp]
  \centering
  \caption{Ideological position (DW-NOMINATE scores) for all members of 109th-115th Congresses by party }
    \begin{tabular}{lcccccccc}
    \toprule
    Congress & Mean  & St. Dev. & Min.  & Max.  & p(25) & p(50) & p(75) & Obs. \\
    \midrule
    \multicolumn{9}{l}{\textit{Republican Representatives}}
\\
    109   & 0.421 & 0.143 & 0.103 & 0.863 & 0.316 & 0.408 & 0.523 & 232 \\
    110   & 0.436 & 0.143 & 0.129 & 0.863 & 0.333 & 0.426 & 0.538 & 202 \\
    111   & 0.456 & 0.145 & 0.133 & 0.913 & 0.348 & 0.441 & 0.555 & 178 \\
    112   & 0.469 & 0.147 & 0.164 & 0.913 & 0.352 & 0.470 & 0.577 & 242 \\
    113   & 0.482 & 0.147 & 0.164 & 0.913 & 0.367 & 0.491 & 0.591 & 234 \\
    114   & 0.480 & 0.149 & 0.164 & 0.829 & 0.362 & 0.490 & 0.600 & 247 \\
    115   & 0.487 & 0.149 & 0.164 & 0.931 & 0.374 & 0.502 & 0.600 & 241 \\
    All   & 0.463 & 0.148 & 0.103 & 0.931 & 0.349 & 0.456 & 0.576 & 1576 \\
        &       &       &       &       &       &       &       &  \\
    \multicolumn{9}{l}{\textit{Democratic Representatives}}
\\
    109   & -0.387 & 0.124 & -0.683 & -0.045 & -0.473 & -0.390 & -0.302 & 202 \\
    110   & -0.369 & 0.132 & -0.683 & -0.045 & -0.463 & -0.379 & -0.282 & 233 \\
    111   & -0.349 & 0.145 & -0.683 & 0.088 & -0.448 & -0.350 & -0.262 & 257 \\
    112   & -0.395 & 0.123 & -0.683 & -0.070 & -0.478 & -0.401 & -0.320 & 193 \\
    113   & -0.383 & 0.115 & -0.683 & -0.104 & -0.460 & -0.390 & -0.306 & 201 \\
    114   & -0.395 & 0.108 & -0.683 & -0.104 & -0.463 & -0.398 & -0.322 & 188 \\
    115   & -0.390 & 0.113 & -0.692 & -0.104 & -0.460 & -0.396 & -0.310 & 194 \\
    All   & -0.379 & 0.126 & -0.692 & 0.088 & -0.464 & -0.389 & -0.301 & 1468 \\
    \bottomrule
    \end{tabular}%
  \label{tab:Nom_House_all}%
\end{table}%

\begin{table}[H]
  \centering
  \caption{Shifts in Democrat vote margin in US House general elections with above/below threshold shares.}
    \resizebox{!}{.45\textheight}{\begin{tabular}{lccccc}
    \toprule
          & USH (D) & USH (D) & USH (D) & USH (D) & USH (D) \\
    VARIABLES & (1)   & (2)   & (3)   & (4)   & (5) \\
    \midrule
          &       &       &       &       &  \\
    Share Close Thresh - Above (5) & -0.235 &       &       &       &  \\
          & (0.152) &       &       &       &  \\
    Share Close Thresh - Below (5) & -0.461 &       &       &       &  \\
          & (0.315) &       &       &       &  \\
    Share Close Thresh - Above (10) &       & -0.263* &       &       &  \\
          &       & (0.148) &       &       &  \\
    Share Close Thresh - Below (10) &       & -0.246 &       &       &  \\
          &       & (0.183) &       &       &  \\
    Share Close Thresh - Above (15) &       &       & -0.258** &       &  \\
          &       &       & (0.125) &       &  \\
    Share Close Thresh - Below (15) &       &       & -0.272* &       &  \\
          &       &       & (0.156) &       &  \\
    Share Close Thresh - Above (20) &       &       &       & -0.279** &  \\
          &       &       &       & (0.114) &  \\
    Share Close Thresh - Below (20) &       &       &       & -0.169 &  \\
          &       &       &       & (0.128) &  \\
    Share Close Thresh - Above (25) &       &       &       &       & -0.276*** \\
          &       &       &       &       & (0.105) \\
    Share Close Thresh - Below (25) &       &       &       &       & -0.139 \\
          &       &       &       &       & (0.121) \\
    Share China Import & 0.016 & 0.017 & 0.017 & 0.017 & 0.017 \\
          & (0.014) & (0.014) & (0.014) & (0.014) & (0.014) \\
    Share White & -0.693*** & -0.695*** & -0.698*** & -0.698*** & -0.700*** \\
          & (0.019) & (0.020) & (0.020) & (0.020) & (0.020) \\
    Share Female (voting age) & 0.971*** & 0.970*** & 0.967*** & 0.966*** & 0.963*** \\
          & (0.179) & (0.179) & (0.179) & (0.179) & (0.179) \\
          &       &       &       &       &  \\
    Observations & 12,053,931 & 12,053,931 & 12,053,931 & 12,053,931 & 12,053,931 \\
    R-squared & 0.420 & 0.420 & 0.420 & 0.421 & 0.421 \\
    \midrule
    \multicolumn{6}{l}{Clustered SE at county x congressional district} \\
    \multicolumn{6}{l}{*** $p<0.01$, ** $p<0.05$, * $p<0.1$} \\
    \end{tabular}%
    }%
  \label{tab:App_Dem_bw_abovebelow}%
\end{table}%


\begin{table}[htbp]
  \centering
  \caption{Shifts in Republican vote margin in US House general elections with above/below threshold shares.}
    \resizebox{!}{.45\textheight}{\begin{tabular}{lccccc}
    \toprule
          & USH (R) & USH (R) & USH (R) & USH (R) & USH (R) \\
    VARIABLES & (1)   & (2)   & (3)   & (4)   & (5) \\
    \midrule
          &       &       &       &       &  \\
    Share Close Thresh - Above (5) & 0.280* &       &       &       &  \\
          & (0.156) &       &       &       &  \\
    Share Close Thresh - Below (5) & 0.445 &       &       &       &  \\
          & (0.308) &       &       &       &  \\
    Share Close Thresh - Above (10) &       & 0.257* &       &       &  \\
          &       & (0.148) &       &       &  \\
    Share Close Thresh - Below (10) &       & 0.266 &       &       &  \\
          &       & (0.185) &       &       &  \\
    Share Close Thresh - Above (15) &       &       & 0.248** &       &  \\
          &       &       & (0.126) &       &  \\
    Share Close Thresh - Below (15) &       &       & 0.292* &       &  \\
          &       &       & (0.158) &       &  \\
    Share Close Thresh - Above (20) &       &       &       & 0.262** &  \\
          &       &       &       & (0.114) &  \\
    Share Close Thresh - Below (20) &       &       &       & 0.186 &  \\
          &       &       &       & (0.131) &  \\
    Share Close Thresh - Above (25) &       &       &       &       & 0.260** \\
          &       &       &       &       & (0.104) \\
    Share Close Thresh - Below (25) &       &       &       &       & 0.156 \\
          &       &       &       &       & (0.124) \\
    Share China Import & -0.018 & -0.018 & -0.019 & -0.019 & -0.019 \\
          & (0.014) & (0.014) & (0.014) & (0.014) & (0.014) \\
    Share White & 0.700*** & 0.702*** & 0.705*** & 0.705*** & 0.707*** \\
          & (0.019) & (0.020) & (0.020) & (0.020) & (0.020) \\
    Share Female (voting age) & -0.974*** & -0.972*** & -0.969*** & -0.969*** & -0.966*** \\
          & (0.183) & (0.183) & (0.183) & (0.183) & (0.183) \\
          &       &       &       &       &  \\
    Observations & 12,053,931 & 12,053,931 & 12,053,931 & 12,053,931 & 12,053,931 \\
    R-squared & 0.435 & 0.435 & 0.435 & 0.435 & 0.435 \\
    \midrule
    \multicolumn{6}{l}{Clustered SE at county x congressional district} \\
    \multicolumn{6}{l}{*** $p<0.01$, ** $p<0.05$, * $p<0.1$} \\
    \end{tabular}%
    }%
  \label{tab:App_Rep_bw_abovebelow}%
\end{table}%


\begin{table}[htbp]
  \centering
  \caption{Shifts in Democrat vote margin in US House general elections with proximity to threshold shares.}
    \begin{tabular}{lccccc}
    \toprule
          & USH (D) & USH (D) & USH (D) & USH (D) & USH (D) \\
    VARIABLES & (1)   & (2)   & (3)   & (4)   & (5) \\
    \midrule
          &       &       &       &       &  \\
    Share Close Thresh (5) & -0.327** &       &       &       &  \\
          & (0.162) &       &       &       &  \\
    Share Close Thresh (10) &       & -0.255* &       &       &  \\
          &       & (0.133) &       &       &  \\
    Share Close Thresh (15) &       &       & -0.264** &       &  \\
          &       &       & (0.118) &       &  \\
    Share Close Thresh (20) &       &       &       & -0.227** &  \\
          &       &       &       & (0.101) &  \\
    Share Close Thresh (25) &       &       &       &       & -0.212** \\
          &       &       &       &       & (0.094) \\
    Share China Import & 0.016 & 0.017 & 0.017 & 0.017 & 0.017 \\
          & (0.014) & (0.014) & (0.014) & (0.014) & (0.014) \\
    Share White  & -0.693*** & -0.695*** & -0.698*** & -0.699*** & -0.700*** \\
          & (0.019) & (0.020) & (0.020) & (0.020) & (0.020) \\
    Share Female (voting age) & 0.972*** & 0.969*** & 0.967*** & 0.967*** & 0.964*** \\
          & (0.179) & (0.179) & (0.179) & (0.179) & (0.179) \\
          &       &       &       &       &  \\
    Observations & 12,053,931 & 12,053,931 & 12,053,931 & 12,053,931 & 12,053,931 \\
    R-squared & 0.420 & 0.420 & 0.420 & 0.420 & 0.421 \\
    \midrule
    \multicolumn{6}{l}{Clustered SE at county x congressional district} \\
    \multicolumn{6}{l}{*** $p<0.01$, ** $p<0.05$, * $p<0.1$} \\
    \end{tabular}%
  \label{tab:App_Dem_bw}%
\end{table}%


\begin{table}[htbp]
  \centering
  \caption{Shifts in Republican vote margin in US House general elections with proximity to threshold shares.}
    \begin{tabular}{lccccc}
    \toprule
          & USH (R) & USH (R) & USH (R) & USH (R) & USH (R) \\
    VARIABLES & (1)   & (2)   & (3)   & (4)   & (5) \\
    \midrule
          &       &       &       &       &  \\
    Share Close Thresh (5) & 0.347** &       &       &       &  \\
          & (0.165) &       &       &       &  \\
    Share Close Thresh (10) &       & 0.261* &       &       &  \\
          &       & (0.135) &       &       &  \\
    Share Close Thresh (15) &       &       & 0.268** &       &  \\
          &       &       & (0.119) &       &  \\
    Share Close Thresh (20) &       &       &       & 0.226** &  \\
          &       &       &       & (0.103) &  \\
    Share Close Thresh (25) &       &       &       &       & 0.212** \\
          &       &       &       &       & (0.095) \\
    Share China Import & -0.018 & -0.018 & -0.019 & -0.019 & -0.019 \\
          & (0.014) & (0.014) & (0.014) & (0.014) & (0.014) \\
    Share White & 0.700*** & 0.702*** & 0.705*** & 0.706*** & 0.707*** \\
          & (0.019) & (0.020) & (0.020) & (0.020) & (0.020) \\
    Share Female (voting age) & -0.975*** & -0.972*** & -0.969*** & -0.969*** & -0.966*** \\
          & (0.183) & (0.183) & (0.183) & (0.183) & (0.184) \\
          &       &       &       &       &  \\
    Observations & 12,053,931 & 12,053,931 & 12,053,931 & 12,053,931 & 12,053,931 \\
    R-squared & 0.435 & 0.435 & 0.435 & 0.435 & 0.435 \\
    \midrule
    \multicolumn{6}{l}{Clustered SE at county x congressional district} \\
    \multicolumn{6}{l}{*** $p<0.01$, ** $p<0.05$, * $p<0.1$} \\
    \end{tabular}%
  \label{tab:App_Rep_bw}%
\end{table}%


\begin{table}[htbp]
  \centering
  \caption{Shifts in Poole-Rosenthal DW-NOMINATE scores for winning candidates in US House general elections with above/below threshold shares.}
      \resizebox{\textwidth}{!}{
    \begin{tabular}{lcccccccccc}
    \toprule
          & \multicolumn{5}{c}{Republican-winning districts} & \multicolumn{5}{c}{Democrat-winning districts} \\
          \cmidrule(lr){2-6} \cmidrule(lr){7-11}
    VARIABLES & (1)   & (2)   & (3)   & (4)   & (5)   & (6)   & (7)   & (8)   & (9)   & (10) \\
    \midrule      &       &       &       &       &       &       &       &       &       &  \\
    
    Share Close Thresh - Above (5) & -0.138 &       &       &       &       & 0.500* &       &       &       &  \\
          & (0.089) &       &       &       &       & (0.259) &       &       &       &  \\
    Share Close Thresh - Below (5) & -0.021 &       &       &       &       & 0.446 &       &       &       &  \\
          & (0.191) &       &       &       &       & (0.300) &       &       &       &  \\
    Share Close Thresh - Above (10) &       & -0.075 &       &       &       &       & 0.428** &       &       &  \\
          &       & (0.088) &       &       &       &       & (0.205) &       &       &  \\
    Share Close Thresh - Below (10) &       & 0.051 &       &       &       &       & 0.451** &       &       &  \\
          &       & (0.141) &       &       &       &       & (0.221) &       &       &  \\
    Share Close Thresh - Above (15) &       &       & -0.134 &       &       &       &       & 0.478*** &       &  \\
          &       &       & (0.082) &       &       &       &       & (0.168) &       &  \\
    Share Close Thresh - Below (15) &       &       & 0.104 &       &       &       &       & 0.465** &       &  \\
          &       &       & (0.117) &       &       &       &       & (0.186) &       &  \\
    Share Close Thresh - Above (20) &       &       &       & -0.114 &       &       &       &       & 0.397*** &  \\
          &       &       &       & (0.075) &       &       &       &       & (0.145) &  \\
    Share Close Thresh - Below (20) &       &       &       & 0.043 &       &       &       &       & 0.487*** &  \\
          &       &       &       & (0.090) &       &       &       &       & (0.170) &  \\
    Share Close Thresh - Above (25) &       &       &       &       & -0.080 &       &       &       &       & 0.361*** \\
          &       &       &       &       & (0.068) &       &       &       &       & (0.129) \\
    Share Close Thresh - Below (25) &       &       &       &       & 0.037 &       &       &       &       & 0.469*** \\
          &       &       &       &       & (0.087) &       &       &       &       & (0.156) \\
    Share China Import & 0.023*** & 0.022*** & 0.023*** & 0.023*** & 0.023*** & 0.015 & 0.014 & 0.014 & 0.013 & 0.013 \\
          & (0.008) & (0.008) & (0.008) & (0.008) & (0.008) & (0.018) & (0.018) & (0.018) & (0.018) & (0.018) \\
    Share White & 0.047* & 0.048* & 0.048* & 0.047* & 0.048* & 0.275*** & 0.280*** & 0.286*** & 0.290*** & 0.293*** \\
          & (0.025) & (0.025) & (0.025) & (0.025) & (0.025) & (0.021) & (0.021) & (0.021) & (0.022) & (0.022) \\
    Share Female (voting age) & 0.147 & 0.148 & 0.146 & 0.144 & 0.145 & -0.412* & -0.410* & -0.409* & -0.408* & -0.406* \\
          & (0.184) & (0.184) & (0.184) & (0.184) & (0.184) & (0.238) & (0.238) & (0.237) & (0.236) & (0.236) \\
          &       &       &       &       &       &       &       &       &       &  \\
    Observations & 6,275,158 & 6,275,158 & 6,275,158 & 6,275,158 & 6,275,158 & 5,867,112 & 5,867,112 & 5,867,112 & 5,867,112 & 5,867,112 \\
    R-squared & 0.052 & 0.052 & 0.053 & 0.053 & 0.052 & 0.225 & 0.226 & 0.228 & 0.229 & 0.231 \\
    \midrule
    \multicolumn{11}{l}{Clustered SE at county x congressional district} \\
    \multicolumn{11}{l}{*** $p<0.01$, ** $p<0.05$, * $p<0.1$} \\
    \end{tabular}%
    }
  \label{tab:App_Nom1_AboveBelow_DR}%
\end{table}%


\begin{table}[htbp]
  \centering
  \caption{Shifts in Poole-Rosenthal DW-NOMINATE scores for winning candidates in US House general elections with proximity to threshold shares.}
      \resizebox{\textwidth}{!}{
    \begin{tabular}{lcccccccccc}
    \toprule
          & \multicolumn{5}{c}{Democrat-winning candidates} & \multicolumn{5}{c}{Republican-winning candidates} \\
          \cmidrule(lr){2-6} \cmidrule(lr){7-11}
    VARIABLES & (1)   & (2)   & (3)   & (4)   & (5)   & (6)   & (7)   & (8)   & (9)   & (10) \\
        \midrule
          &       &       &       &       &       &       &       &       &       &  \\
    Share Close Thresh (5) & 0.469** &       &       &       &       & -0.104 &       &       &       &  \\
          & (0.215) &       &       &       &       & (0.094) &       &       &       &  \\
    Share Close Thresh (10) &       & 0.441*** &       &       &       &       & -0.026 &       &       &  \\
          &       & (0.159) &       &       &       &       & (0.089) &       &       &  \\
    Share Close Thresh (15) &       &       & 0.471*** &       &       &       &       & -0.035 &       &  \\
          &       &       & (0.145) &       &       &       &       & (0.080) &       &  \\
    Share Close Thresh (20) &       &       &       & 0.441*** &       &       &       &       & -0.043 &  \\
          &       &       &       & (0.125) &       &       &       &       & (0.068) &  \\
    Share Close Thresh (25) &       &       &       &       & 0.413*** &       &       &       &       & -0.027 \\
          &       &       &       &       & (0.113) &       &       &       &       & (0.064) \\
    Share China Import & 0.015 & 0.014 & 0.014 & 0.013 & 0.013 & 0.023*** & 0.023*** & 0.023*** & 0.023*** & 0.023*** \\
          & (0.018) & (0.018) & (0.018) & (0.018) & (0.018) & (0.008) & (0.008) & (0.008) & (0.008) & (0.008) \\
    Share White & 0.275*** & 0.280*** & 0.286*** & 0.289*** & 0.293*** & 0.047* & 0.048* & 0.047* & 0.047* & 0.047* \\
          & (0.021) & (0.021) & (0.021) & (0.021) & (0.022) & (0.025) & (0.025) & (0.025) & (0.025) & (0.025) \\
    Share Female (voting age) & -0.412* & -0.410* & -0.409* & -0.408* & -0.405* & 0.146 & 0.147 & 0.146 & 0.145 & 0.145 \\
          & (0.238) & (0.238) & (0.237) & (0.236) & (0.236) & (0.184) & (0.184) & (0.184) & (0.184) & (0.184) \\
          &       &       &       &       &       &       &       &       &       &  \\
    Observations & 5,867,112 & 5,867,112 & 5,867,112 & 5,867,112 & 5,867,112 & 6,275,158 & 6,275,158 & 6,275,158 & 6,275,158 & 6,275,158 \\
    R-squared & 0.225 & 0.226 & 0.228 & 0.229 & 0.230 & 0.052 & 0.052 & 0.052 & 0.052 & 0.052 \\
    \midrule
    \multicolumn{11}{l}{Clustered SE at county x congressional district} \\
    \multicolumn{11}{l}{*** $p<0.01$, ** $p<0.05$, * $p<0.1$} \\
    \end{tabular}%
    }
  \label{tab:App_Nom1_allshare_DR}%
\end{table}%


\begin{table}[htbp]
  \centering
  \caption{Shifts in Poole-Rosenthal DW-NOMINATE scores for all candidates in US House general elections with proximity to threshold shares.}
    \begin{tabular}{lccccc}
    \toprule
    VARIABLES & (1)   & (2)   & (3)   & (4)   & (5) \\
    \midrule
          &       &       &       &       &  \\
    Share Close Thresh (5) & 0.639* &       &       &       &  \\
          & (0.347) &       &       &       &  \\
    Share Close Thresh (10) &       & 0.577** &       &       &  \\
          &       & (0.287) &       &       &  \\
    Share Close Thresh (15) &       &       & 0.509** &       &  \\
          &       &       & (0.258) &       &  \\
    Share Close Thresh (20) &       &       &       & 0.406* &  \\
          &       &       &       & (0.225) &  \\
    Share Close Thresh (25) &       &       &       &       & 0.392* \\
          &       &       &       &       & (0.206) \\
    Share China Import & 0.031 & 0.030 & 0.030 & 0.029 & 0.029 \\
          & (0.029) & (0.029) & (0.029) & (0.029) & (0.029) \\
    Share White & 1.093*** & 1.099*** & 1.103*** & 1.103*** & 1.107*** \\
          & (0.048) & (0.048) & (0.048) & (0.048) & (0.049) \\
    Share Female (voting age) & -1.047** & -1.040** & -1.037** & -1.037** & -1.032** \\
          & (0.434) & (0.434) & (0.434) & (0.434) & (0.434) \\
          &       &       &       &       &  \\
    Observations & 12,146,679 & 12,146,679 & 12,146,679 & 12,146,679 & 12,146,679 \\
    R-squared & 0.296 & 0.296 & 0.296 & 0.296 & 0.296 \\
    \bottomrule
 \multicolumn{6}{l}{Clustered SE at county x congressional district} \\
    \multicolumn{6}{l}{*** $p<0.01$, ** $p<0.05$, * $p<0.1$} \\
    \end{tabular}%
  \label{tab:app_nom1_allcand_allshare}%
\end{table}%

\FloatBarrier
\subsection{Controlling for the Gerrymandering of Congressional Districts}\label{sec:appendix_gerry}

\begin{table}[H]
  \centering
  \caption{Shifts in vote margins for Democrat and Republican candidates in US House general elections with 15 credit score point bandwidth.}
    \begin{tabular}{lcccc}
    \toprule
          & \multicolumn{2}{c}{Republican votes} & \multicolumn{2}{c}{Democrat Votes} \\
          \cmidrule(lr){2-3} \cmidrule(lr){4-5}
    VARIABLES & (1)   & (2)   & (3)   & (4) \\
    \midrule
          &       &       &       &  \\
    Share Close Thresh &       & 0.541*** &       & -0.497*** \\
          &       & (0.163) &       & (0.170) \\
    Share Close Thresh (Above) & 0.357** &       & -0.301 &  \\
          & (0.179) &       & (0.187) &  \\
    Share Close Thresh (Below) & 0.766*** &       & -0.738*** &  \\
          & (0.218) &       & (0.225) &  \\
    Share China Import & 0.006 & 0.006 & -0.016 & -0.016 \\
          & (0.016) & (0.016) & (0.016) & (0.016) \\
    Share White & 0.807*** & 0.806*** & -0.787*** & -0.786*** \\
          & (0.027) & (0.027) & (0.028) & (0.029) \\
    Share Female (voting age) & -0.588** & -0.587** & 0.699*** & 0.698*** \\
          & (0.259) & (0.258) & (0.259) & (0.259) \\
          &       &       &       &  \\
    Observations & 4,180,761 & 4,180,761 & 4,180,761 & 4,180,761 \\
    R-squared & 0.503 & 0.503 & 0.478 & 0.477 \\
    \midrule
    \multicolumn{5}{l}{Clustered SE at county x congressional district} \\
    \multicolumn{5}{l}{*** $p<0.01$, ** $p<0.05$, * $p<0.1$}\\
    \end{tabular}%
  \label{tab:app_GerryM_votes}%
  \caption*{\footnotesize The sample is restricted to election years 2012, 2014, and 2016 to exclude districts subject to potentially strategic redistricting after the 2010 census.}
\end{table}%


\begin{table}[htbp]
  \centering
  \caption{Shifts in Poole-Rosenthal DW-NOMINATE scores for Democrat, Republican, and all candidates in US House general elections with 15 credit score point bandwidth.}
    \begin{tabular}{lcccccc}
    \toprule
          & \multicolumn{2}{c}{Republican votes} & \multicolumn{2}{c}{Democrat Votes} & \multicolumn{2}{c}{All Candidates} \\
          \cmidrule(lr){2-3} \cmidrule(lr){4-5} \cmidrule(lr){6-7}
    VARIABLES & (1)   & (2)   & (3)   & (4)   & (5)   & (6) \\
    \midrule
          &       &       &       &       &       &  \\
    Share Close Thresh &       & -0.167 &       & 0.462* &       & 0.319 \\
          &       & (0.104) &       & (0.269) &       & (0.390) \\
    Share Close Thresh (Above) & -0.230* &       & 0.446 &       & 0.253 &  \\
          & (0.127) &       & (0.326) &       & (0.431) &  \\
    Share Close Thresh (Below) & -0.085 &       & 0.479 &       & 0.399 &  \\
          & (0.147) &       & (0.335) &       & (0.491) &  \\
    Share China Import & 0.028** & 0.028** & -0.039 & -0.039 & 0.098** & 0.098** \\
          & (0.012) & (0.012) & (0.032) & (0.032) & (0.038) & (0.038) \\
    Share White & 0.034 & 0.034 & 0.262*** & 0.262*** & 1.200*** & 1.200*** \\
          & (0.028) & (0.028) & (0.037) & (0.037) & (0.066) & (0.066) \\
    Share Female (voting age) & 0.291 & 0.292 & -0.390 & -0.390 & -0.132 & -0.132 \\
          & (0.234) & (0.234) & (0.352) & (0.352) & (0.590) & (0.590) \\
          &       &       &       &       &       &  \\
    Observations & 2,307,950 & 2,307,950 & 1,886,554 & 1,886,554 & 4,194,504 & 4,194,504 \\
    R-squared & 0.020 & 0.020 & 0.134 & 0.134 & 0.329 & 0.329 \\
    \midrule
    \multicolumn{7}{l}{Clustered SE at county x congressional district} \\
    \multicolumn{7}{l}{*** $p<0.01$, ** $p<0.05$, * $p<0.1$} \\
    \end{tabular}%
  \label{tab:app_GerryM_nom1}%
    \caption*{\footnotesize The sample is restricted to election years 2012, 2014, and 2016 to exclude districts subject to potentially strategic redistricting after the 2010 census.}
\end{table}%

\end{document}